\documentclass{aa}
\usepackage{graphics}

\def\Teff{${\rm T_{eff}}$\,}
\def\logg{$\log g$\,}
\def\vt{$v_{\rm t}$\,}

\begin{document}

\thesaurus{08.02.4; 08.12.1; 10.07.3; 11.13.1}

\title{Age-metallicity relation and Chemical evolution of the LMC from
 UVES spectra of Globular Cluster giants \thanks{ based on observations made 
 at the ESO Telescopes in Chile}}

\author{V. Hill \inst{1} \and 
P. Fran\c cois \inst{1} \and
M. Spite \inst{2} \and  
F. Primas \inst{1} \and
F. Spite \inst{2}}
\offprints{V.Hill}

\institute{ 
European Southern Observatory, D-85748 Garching, Germany  \and
Observatoire de Paris-Meudon, F-92125 Meudon Cedex, France 
}

\date{ }

\authorrunning{V. Hill et al.}

\titlerunning{Chemical evolution of the LMC from abundances in Globular Clusters}

\maketitle

\begin{abstract}
We report on the first high-resolution spectroscopy of 10 giants in
LMC Globular Clusters {\em in a wide age range}, obtained with the newly
commissioned spectrograph UVES at VLT UT2. These observations are used
to derive oxygen and iron content of these clusters, and the
abundances are then used to cast a more precise view, not only on the
age-metallicity relation in the LMC, but also on the chemical
evolution of this dwarf irregular galaxy.
\end{abstract}


\section{Introduction}

The Magellanic Clouds (MCs) possess a large population of stellar 
clusters of a whole range of ages, including 13 {\it bona fide} old 
Globular Clusters in the LMC and only one somewhat younger counterpart 
in the SMC. The photometry of these clusters has been extensively 
studied over the years, and especially since HST started producing 
high quality CMDs reaching the very densest central part of the 
clusters (Olsen et al. \cite{olsen 98}, Johnson et
al. \cite{johnson 99}).

The opportunities offered by these clusters are twofold: they provide
laboratories to study stellar evolution in a range of different 
conditions (the young and yet metal-poor MC clusters have no 
Galactic counterparts and serve as a test of stellar evolution models at 
low metallicities), and give us the opportunity to study the formation 
and evolution of magellanic-type galaxies (determination of the 
age-metallicity relation, etc\ldots). 
It is with this second aspect that we will be dealing in this paper.

In particular, the oldest component of these stellar clusters (the 
{\it  bona fide} old metal-poor globular clusters) can serve as tracers of
early epochs of the evolution (halo phases) in several ways: 
precise determination of ages of these old globular clusters
yield estimates of the characteristic time-scales of halo formation;
information on the chemical evolution of the MCs at early
stages of their evolution is kept in the atmosphere of the cluster
stars; kinematics of the clusters can be used as a tracer of halo
structure. 

While kinematics issues were addressed at length already some 
time ago (e.g. Schommer et al. \cite{schommer 92}), topics in need of precise 
metallicities were so far quite
tentative: because of the faintness of their members,
chemical composition of old and intermediate age clusters
were not known with a very good accuracy (low-resolution Ca\,II IR
triplet, photometric estimates). We are now for the first time, thanks
to VLT and its efficient spectrograph UVES,  
in a position to improve dramatically this knowledge
with detailed analysis of individual cluster giants fainter than 16 magnitudes.

As part of UVES {\em Science Verification}, a sample of 10 giants in 4
clusters in a very wide age-range (0.1 to 15\,Gyrs) were observed at high spectral
resolution (R$\simeq$40000). In this paper, we present the analysis of this
data, derive oxygen and iron content of the four clusters, and use
them to further constrain the age-metallicity relation in the LMC. We
also present, for the first time, the evolution of the
[O/Fe] along time in the LMC, and briefly discuss the implications on
the processes driving the chemical evolution of this dwarf irregular
galaxy.
A future paper (in preparation) will deal with the full analysis of the 20
elements measured in these giants and the full implication on the
chemical evolution of the LMC.

\section{Observations}

\begin{table}
\caption{Log book of the observations}\label{T-log}
\begin{flushleft}
\begin{tabular}{lcccc}
\noalign{\smallskip}
\hline
\noalign{\smallskip}
Star  & MJD & slit & Exp.   & S/N \\
      &     &$\arcsec$ &h.m &     \\
\hline
\noalign{\smallskip}
NGC\,1866 B444     & 51584.10565 &1.2 & 1h30 &75 \\
NGC\,1866 B1653    & 51586.04048 &1.2 & 1h00 &85 \\
NGC\,1866 B867     & 51591.03805 &1.0 & 1h00 &  \\
                  & 51591.08042 &1.0 & 1h00 &40 \\
NGC\,1978 LE8      & 51590.03804 &1.2 & 1h15 & \\
                  & 51590.09094 &1.2 & 1h15 &70 \\
NGC\,1978 LE9      & 51589.03189 &1.0 & 1h15 & \\
                  & 51589.08487 &1.0 & 1h15 &70 \\
ESO\,121-SCO3  M313     & 51585.04156 &1.0 & 1h30 & \\
                  & 51585.10510 &1.0 & 1h30 &45 \\
ESO\,121-SCO3  M167     & 51587.03766 &1.0 & 1h15 & \\
                  & 51588.10597 &1.0 & 1h15 &60 \\
NGC\,2210 B4364$^*$& 51586.08746 &1.0 & 1h40 &55 \\
NGC\,2210 B110$^*$ & 51586.08746 &1.0 & 1h40 &30 \\
NGC\,2210 B4793    & 51588.04009 &1.0 & 1h30 &60 \\
\noalign{\smallskip}
\hline
\end{tabular}
Identifications for the clusters: NGC\,1866: Brocato et al. \cite{brocato 89};
NGC\,1978: Lloyd Evans \cite{lloyd 80};
ESO\,121-SC03 Mateo et al \cite{mateo 86}; 
NGC\,2210 Brocato et al. \cite{brocato 96}.\\
$^*$ the two stars were observed on the same slit (12 arcsec long). 
\end{flushleft}
\end{table}

The observations were performed from the 10 to the 18th of February 2000, 
as part of the VLT UT2 {\em Science Verification} of UVES, the UV and 
Visual Echelle Spectrograph recently commissioned on VLT (D'Odorico et al. 
\cite{dodorico 00}). The setting used 
(RED arm, $\lambda_{\rm central}$ 5800\AA) 
provided a wavelength coverage 4800-6800\AA. The entrance slit was
adjusted to fit the seeing conditions, between 1 and 1.2$\arcsec$, which
translate respectively into a resolving power of R$\sim$45000 to
38000.  
The table \ref{T-log} gives a short logbook of the observations, together 
with the achieved S/N (per 0.03\AA\, pixel, at 6000\AA).

The spectra were reduced using the UVES context within MIDAS, which 
performs bias subtraction (object and flat-field), inter-order 
background subtraction (object and flat-field), optimal extraction of 
the object (above the sky, rejecting cosmic hits), division by a 
flat-field frame extracted with the same weighted-profile as the 
object, wavelength calibration and re-binning to a constant step and 
merging of all overlapping orders.  The spectra were then corrected 
for barycentric radial velocity and coadded, and finally normalized.

\section{Abundance analysis}

A more detailed discussion will be given in
a paper (Hill et al., in preparation) dealing with the full abundance analysis
of the 20 elements detected in these giants.
We give however here a short summary of the methods used to determine
the iron and oxygen abundances.

\subsection{Atmospheric parameters}

The effective temperature \Teff and gravity \logg of each star were 
determined in a two-step process: \\
1- The colour and absolute magnitudes of 
the cluster stars were used to derive the effective temperatures using the 
Alonso et al. (\cite{alonso 99}) calibrations for giants, and the gravity 
using the simple relation between gravity and absolute luminosity (once 
\Teff and masses are known). The colour information, together with the 
derived temperatures and gravities of each program star are given in the 
Table \ref{T-param} (columns 1 to 7). For each cluster, the table also  
lists the reddening, distance modulus, metallicity and masses
(isochrones Bertelli et al. \cite{bertelli 94}) adopted for the
computation of the \Teff and \logg.\\
2- In a second step, we used spectroscopic indices to further constrain the 
atmospheric parameters:\\
$\bullet$ \Teff was constrained requiring the excitation equilibrium of Fe\,I to be fulfilled.\\
$\bullet$ \logg was determined from the ionization balance of Fe\,I and II 
(and Ti when available)\\
$\bullet$ the microturbulence velocity (\vt) was determined asking the Fe\,I 
abundance to be independent of the line strength.

The temperatures and gravities deduced in that way (columns 8-10 of 
Table \ref{T-param}) were very close to the ones derived from 
photometry and were adopted for the rest of the analysis.  The only 
counter-example is NGC1978\,LE9, for which the ionization balance 
requires a gravity 0.5\,dex {\em lower} than predicted.  This 
behaviour is well known in very cool galactic metal-poor giants (field 
and Globular Clusters), where it is interpreted as a signature of NLTE 
effects.

The associated uncertainties are expected to
be of the order of 

$\Delta {\rm T_{eff}} = \pm 150K$  
$\Delta \log g = \pm 0.2$ 
$\Delta v_{\rm t} = \pm 0.2 {\rm km \: s^{-1}~}$.
Also listed in Table \ref{T-param} are the measured 
heliocentric radial velocity for each star. 

\begin{table*}
\caption{Photometry, atmospheric parameters and measured abundances
for the program stars.}\label{T-param}
\begin{flushleft}
\begin{tabular}{lccc@{ }c@{ }c@{ }c@{ }c@{ }c@{ }ccccc}
\noalign{\smallskip}
\hline
\noalign{\smallskip}
\noalign{NGC\,1866 E$(B-V)$=0.06  $(M-m)$=18.6 [Fe/H]=-0.4 Mass=5(4)$M_{\sun}$} 
Star & $V$ & $(B-V)$ &  & Ref.& \Teff$_{\rm phot}$ & \logg$_{\rm phot}$ & 
\Teff & \logg & \vt & Vr$_{\rm helio}$ & [Fe/H] & [O/Fe] & [Al/Fe]\\
B444  &15.49 &1.51 &     & 1 & 4021 & 1.0 &4020 & 1.0 &1.9 & 300.5
&-0.45 &  0.03 & $-$0.23\\ 
B1653 &15.01 &1.49 &     & 1 & 4048 & 0.8 &4050  & 0.8 &2.0 & 300.7
&-0.48 &  0.06 & $-$0.27\\
B867  &16.60 &1.14 &     & 1 & 4550 & 1.7 &4550 & 1.9 &1.8 & 298.2
&-0.56 &  0.14 & $-$0.25\\
\hline
\noalign{NGC1978 E$(B-V)$=0.08  $(M-m)$=18.5  [Fe/H]=-0.4 Mass=2.0$M_{\sun}$} 
Star & $V$ & $(B-V)$ & $(J-K)$ & Ref.& \Teff$_{\rm phot}$ &
\logg$_{\rm phot}$ & 
\Teff & \logg & \vt & Vr$_{\rm helio}$& [Fe/H] & [O/Fe] & [Al/Fe]\\
LE8   &16.71 &1.66 & 1.09& 2,3 & 3860/4000 & 0.8 &3860  &0.7  &1.9 & 292.2
&-1.10 &  0.45 &  \hspace{1ex}0.28\\
LE9   &16.85 &1.77 & 0.94& 2,3 & 3734/3800 & 0.8 &3750  &0.3  &1.9 & 294.7
&-0.82 &  0.30 & $-$0.07\\
\hline
\noalign{ESO\,121-SCO3 E$(B-V)$=0.03   $(M-m)$=18.5  [Fe/H]=-1   Mass=1.0$M_{\sun}$} 
Star & $V$ & $(B-V)$ &  & Ref.& \Teff$_{\rm phot}$ & \logg$_{\rm phot}$ & 
\Teff & \logg & \vt & Vr$_{\rm helio}$& [Fe/H] & [O/Fe] & [Al/Fe]\\
M313  &17.05 &1.29 &     & 4 & 4224 & 1.1 &4220 &1.1 &1.7 & 310.5
&-0.89 &  0.15 & $-$0.43\\
M167  &16.74 &1.47 &     & 4 & 3999 & 0.9 &4000 &0.9 &1.7 & 313.9
&-0.93 &  0.15 & $-$0.33\\
\hline
\noalign{NGC2210  E$(B-V)$=0.06 $(M-m)$=18.4  [Fe/H]=-2.0 Mass=1.0$M_{\sun}$} 
Star & $V$ & & $(V-I)$ & Ref.& \Teff$_{\rm phot}$ & \logg$_{\rm phot}$ & 
\Teff & \logg & \vt & Vr$_{\rm helio}$& [Fe/H] & [O/Fe] & [Al/Fe]\\
B4364 &16.22 &     &1.39 & 5 & 4261 & 0.7 &4260 &0.7 &1.8 & 342.2
&-1.78 &  0.02 &  \hspace{1ex}0.47\\
B110  &16.89 &     &1.21 & 5 & 4570 & 1.2 &4570 &1.0 &1.8 & 344.2
&-1.81 &  0.21 & $-$0.14\\
B4793 &16.12 &     &1.41 & 5 & 4231 & 0.6 &4230 &0.7 &1.8 & 338.8
&-1.68 &  0.19 &  \hspace{1ex}0.00\\
\noalign{\smallskip}
\hline
\end{tabular}
References:1: Brocato et al. \cite{brocato 89}
2: Will et al. \cite{will 95}
3: Ferraro et al \cite{ferraro 95}
4: Mateo et al. \cite{mateo 86}
5: Brocato et al. \cite{brocato 96}\\
Note:{\it
The measured $(V-I)_{\rm Cousins}$ were converted into $(V-I)_{\rm Johnson}$ 
(suited for the Alonso et al. (\cite{alonso 99}) temperature calibration)
using the relation $(V-I)_{\rm Johnson}=1.273\times (V-I)_{\rm
Cousins}-0.05$.} 
\end{flushleft}
\end{table*}

\subsection{Abundance determinations}

The abundance calculations were made using model atmospheres from the 
MARCS suite: the models of Plez (\cite{plez 98}, a grid especially 
computed for cool giants in the metallicity range 
[Fe/H]=$-$1 to +0.3\,dex) were used for the three more metal rich clusters, 
and Gustafsson (\cite{gustafsson 75}), for the more metal poor 
cluster NGC\,2210.
 
\begin{figure}
\center{
\resizebox{7.3cm}{!}{\includegraphics{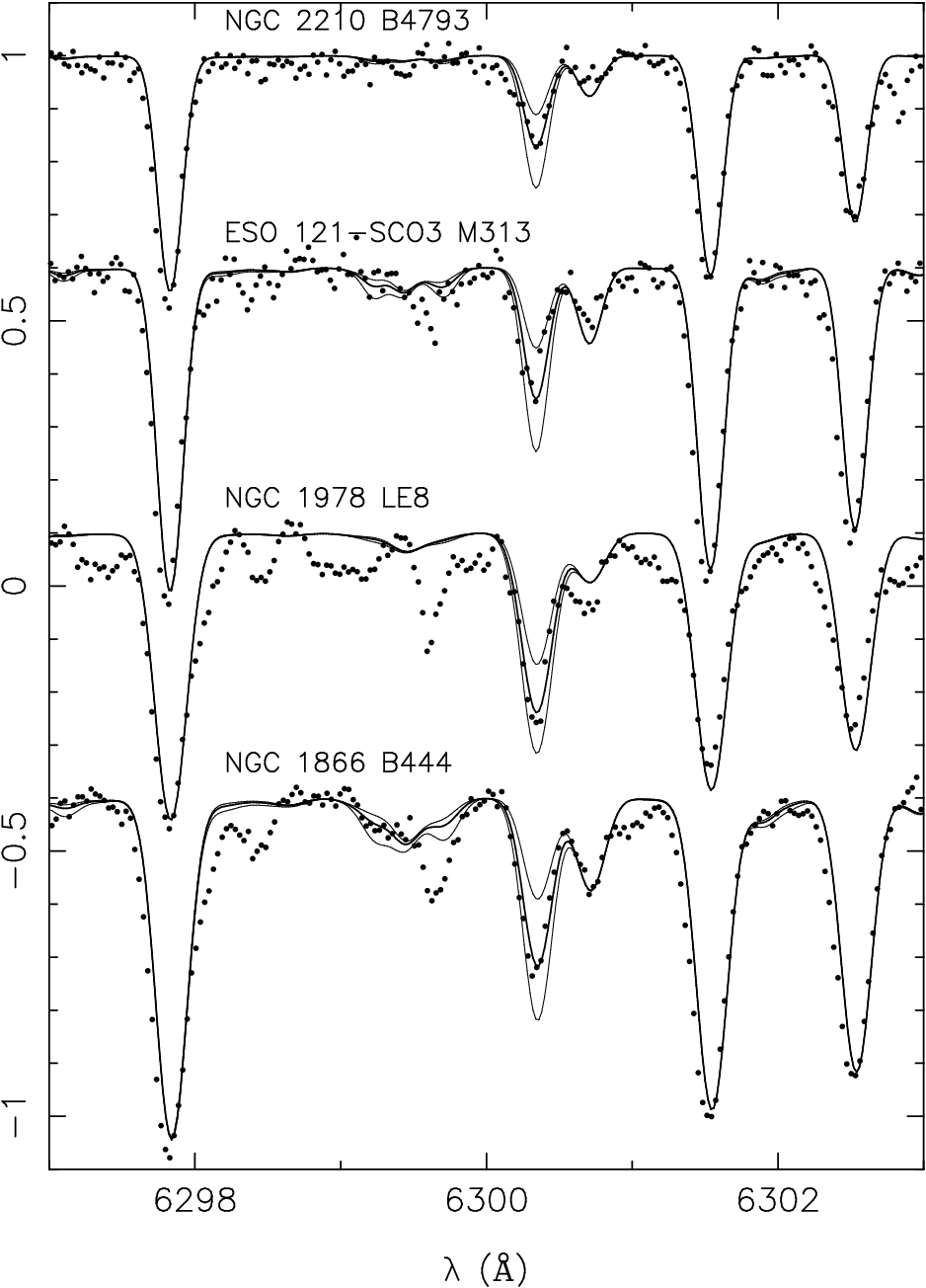}}}
\caption{Examples of oxygen line synthesis. Three synthesis are
shown in each case, computed with the best fitting [O/H] value and 
$\pm$0.2\,dex. The spectra were shifted vertically for 
lisibility purposes.}\label{F-ox}
\end{figure}

The abundance of iron was determined using the equivalent width 
of up to 70 lines of Fe\,I and up to 10\,Fe\,II lines. Aluminum lines
(Al\,I 6696.03\AA\, and 6698.67\AA) were also measured to check for
possible deep-mixing effects (cf. section \ref{mixing}).  
Oxygen abundances, on the other hand, were determined from the forbidden
[OI] 6300\AA\, line, by comparing the line directly to synthetic
spectra. The molecular dissociative equilibrium was taken
into account while computing the synthesis,
including CO, CN, C2, and TiO molecules. 
For  the two younger clusters (NGC\,1866 and NGC\,1978), a telluric 
line was contaminating  the [OI] 6300\AA\, line, and was removed using the
spectra of extremely metal-poor turnoff stars observed during the same
nights.
Examples of the [OI] fits are displayed in Fig.\ref{F-ox}, and the
measured abundances obtained for Fe, O and Al are displayed in
Table \ref{T-param} col. 12-14.

\section{Results}
\subsection{Cluster metallicities}

Both  ESO\,121-SC03 and NGC\,1866 are found to have metallicities
similar to what was previously assumed in the literature. 
ESO\,121-SC03 has a mean iron abundance [Fe/H]=$-0.91\pm 0.16$ (2 stars) 
compared to $-0.93\pm 0.2$ in O91 (Ca IR triplet) and $-0.9$ in  
Mateo et al. (\cite{mateo 86}, BV photometry),
while NGC\,1866  has a mean iron abundance [Fe/H]=$-0.50\pm 0.1$ (3 stars), 
in very good agreement with  Hilker et al. (\cite{hilker 95}) who derived 
[Fe/H]=$-$0.46  using Str\"omgren photometry 
(there are no previous spectroscopic determinations of metallicity of NGC\,1866).

On the other hand, two clusters are found to have a metal-content significantly
different from the previously assumed values:

{\it NGC\,2210:} The mean iron abundance of the 3 giants in this cluster is
[Fe/H]=$-1.75\pm 0.1$, whereas Olszewski et
al. (\cite{olszewski 91}; hereafter O91) derived $-1.97\pm 0.2$ from 4
similar giants. These two values taken at face are only marginally
consistent, but taking into account that the O91 metallicity indicator
are calcium lines calibrated to [Fe/H] using galactic globular clusters as
comparison, this difference might even vanish completely. In effect,
as it is discussed in section \ref{oxygen}  and appears clearly in
Table\ref{T-param}, the [O/Fe] ratio in NGC\,2210 is lower than in
galactic globular clusters of the same metallicity. In fact, as will
be developed in the main paper of this series (in preparation),
all $\alpha$-elements, including calcium, are also less abundant than 
in the galactic clusters. Hence
using the Ca\,II IR triplet would lead to systematically low [Fe/H]
estimates, by $\sim$0.2\,dex.

\begin{figure}
\resizebox{\hsize}{!}{\includegraphics{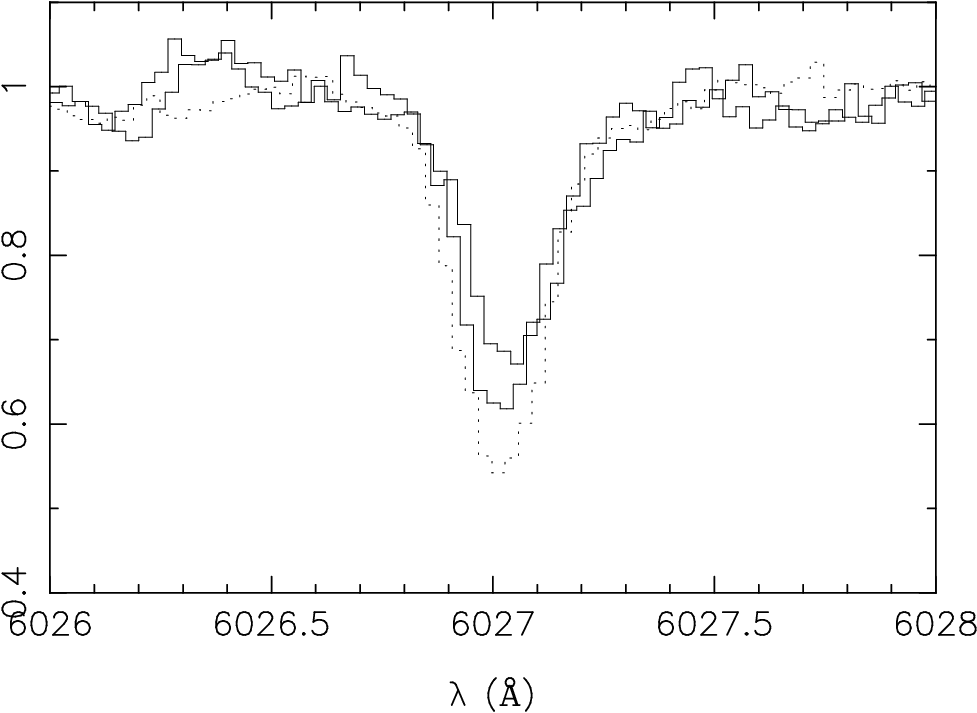}\includegraphics{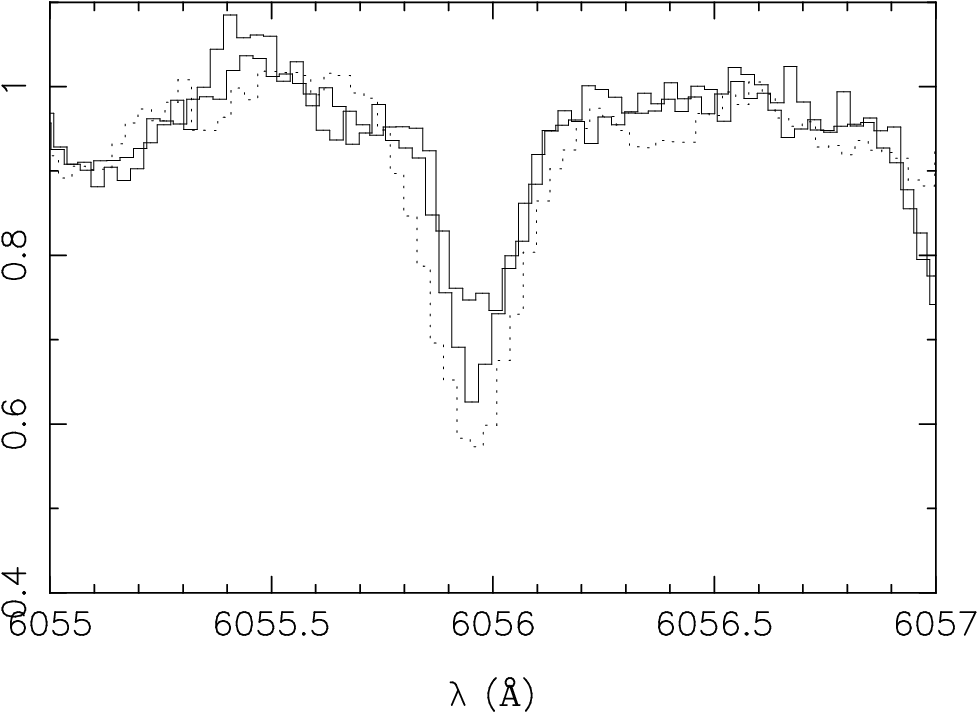}}
\caption{Fe\,I lines of NGC\,1978 LE8 and LE9 (upper and lower solid
lines) compared to the 200-300\,K hotter star NGC\,1866 1653 (dotted
line).}\label{F-1866-1978}   
\end{figure}

{\it NGC\,1978:} The mean iron abundance derived from the 2
giants in NGC\,1978 is [Fe/H]=$-0.96\pm 0.2$, whereas O91 derived
$-0.42\pm 0.2$ from the two {\it same} stars. 
It has to be emphasized that these stars are the coolest giants of our
sample, and the analysis of such objects is always difficult, due to
the presence of molecular features, and to the more uncertain
stellar parameter determination (\Teff and gravity). However, a very
simple test can be performed: in Fig.\ref{F-1866-1978}, we overplotted
Fe\,I lines of NGC\,1978\,LE8 and LE9 to those of NGC\,1866\,B1653. B1653 is
at least 200\,K hotter than the NGC\,1978 giants, since it does not show
any TiO bands, whereas LE8 and LE9 do (dissociation of TiO is too
efficient for \Teff hotter than 3900-4000K), and however, the Fe\,I
lines of B1653 are larger than the ones of LE8 or LE9. NGC\,1978
therefore {\it has to be} significantly more metal poor than
$-0.5$\,dex.

Another interesting issue is that NGC\,1978 has a high ellipticity and
a broadened red giant branch and clump, that could be due to the
merging of two individual clusters. Alcaino et al. (\cite{alcaino 99})
proposes that the two clusters should have similar ages (explaining the
sharp turnoff region), but different metallicities (differing by
$\sim$0.2\,dex) depending on the location in the cluster. The two giants
studied here are indeed located in the south-east of the cluster,
where the most metal-poor population is expected. It would be
extremely interesting to study systematically the abundances of
individual stars as a function of location in the cluster and in the CMD.

\begin{figure}
\resizebox{\hsize}{!}{\includegraphics{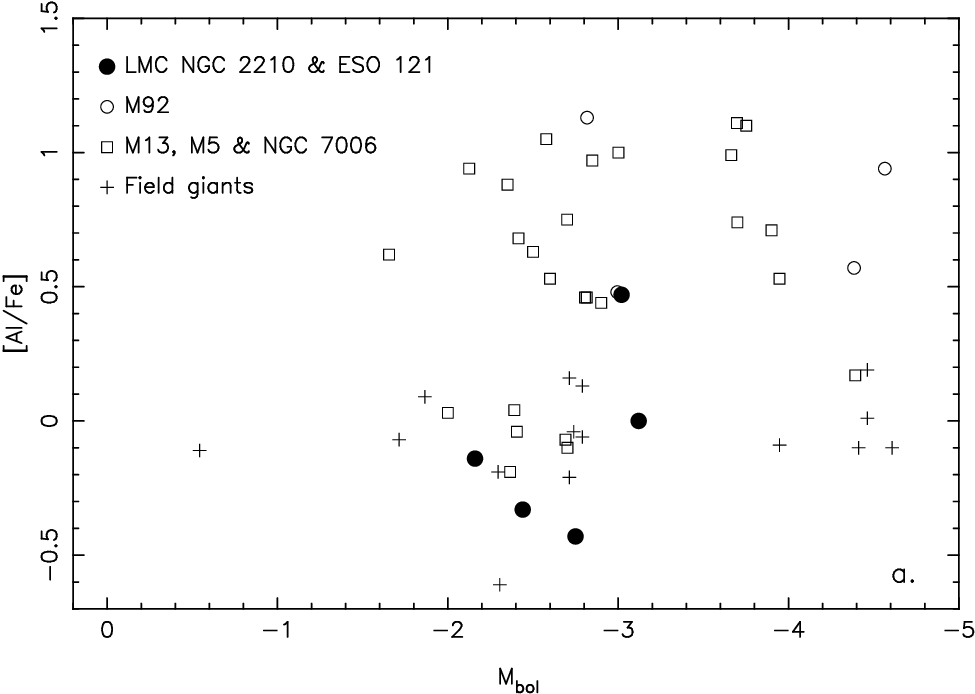}}
\resizebox{\hsize}{!}{\includegraphics{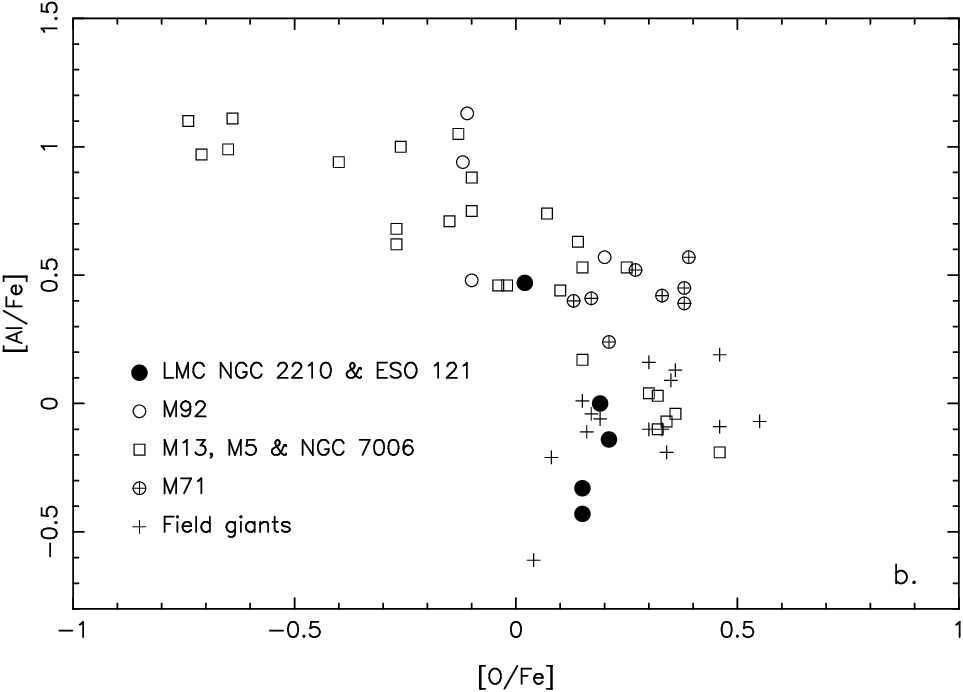}}
\caption{[Al/Fe] abundances for the stars in NGC\,2210 and ESO\,121-SC03 are
plotted, against bolometric magnitude in panel a., and as a function
of [O/Fe] in panel b.  
Overimposed are various samples of red giants in Globular clusters and
in the field of our Galaxy (see text for references), 
illustrating the difference of mixing
experienced by field and cluster stars. With one
exception (NGC2210\,B4364), the LMC
old cluster giants clearly follow the galactic field behaviour.
}\label{F-mix}  
\end{figure}

\subsection{Oxygen abundance and deep mixing effects}\label{mixing}

Oxygen abundances in evolved red giant stars, can be
affected by internal mixing phenomena bringing internally-processed
material to the surface of the star. The oxygen abundance observed at the
surface of the star in this case is not the same as the one the
star was born with.
Before using [O/Fe] ratios for chemical evolution purposes, we
therefore checked the status of mixing in our stars.

In the CNO cycle of hydrogen burning, the final oxygen abundance is
very little affected, the cycle basically turning C into N. So that
giants experiencing ``normal'' shallow mixing do not show depleted nor
enhanced oxygen. However, other cycles involving oxygen, such as the
Ne-Na and the Mg-Al cycles may occur at hotter temperature and hence deeper
inside the star. If mixing mechanisms reach deep enough within
the star, the products of such cycles can be brought up to the
surface, where depleted O and Mg, and enhanced Na and Al can be observed.
A very good tracer of deep mixing are the anti-correlations O-Na, Mg-Al,
or the even more spectacular O-Al.
In our Galaxy, giants in globular clusters and similar giants in the
field do not behave similarly: while O-Na O-Al anti-correlations are
commonly observed in globular clusters of various metallicities
(e.g. Shetrone \cite{shetrone 96}, Kraft et al. \cite{kraft 98}), field
stars of comparable evolutionary stage and metallicities show no such
trends. This difference is generally attributed to environment effects
within the clusters.

In Fig.\ref{F-mix}, we have plotted the data points of the 5 giants
in the 2 older LMC clusters (ESO\,121-SC03 and NGC\,2210)
together with data for
classical galactic globular clusters of three different
metallicities covering the same metallicity range as the LMC clusters. 
M92 has a [Fe/H]=$-2.2$, 
M13, M5 and NGC7006 have [Fe/H] from $-1.$ to $-1.5$, and
M71 is more metal-rich with [Fe/H]=$-0.8$.
All data (galactic clusters and field) are from
Shetrone (\cite{shetrone 96}), except NGC7006 which is from Kraft et
al. (\cite{kraft 98}).
The upper panel shows how the Al overabundances are linked to the
evolutionary stage of the giants (where M$_{\rm bol}$ traces evolution),
and in the lower panel is displayed the O-Al anti-correlation.
Also displayed are the galactic field giants, which clearly do not
follow the same trend, although spanning the same M$_{\rm bol}$ range.

Out of our five LMC old cluster giants, however, only one 
(NGC\,2210\,B4364) has a high [Al/Fe] ratio, while the four others 
have [Al/Fe] ratios similar or even slightly lower than those of the 
field giants of comparable bolometric magnitudes, suggesting that 
these four giants {\it have not experienced major deep mixing events}.  
The anti-correlation between [Al/Fe] and [O/Fe] is also not observed 
as such (the LMC sample is far too small), but as expected, the star 
which is Al-rich is also O-poor: B4364 has most probably dredged up 
material processed through Ne-Na and Mg-Al cycles.

From panel b. of Fig.\ref{F-mix}, it is striking that not only are
the [Al/Fe] ratios low compared to galactic globular cluster stars,
but that the LMC globular cluster giants seem to define a
lower-envelope to the anti-correlation: [O/Fe] seems to be
systematically lower in the LMC giants.
Galactic cluster giants which have [Al/Fe] close to zero (where the 
deep mixing mechanism is not acting), have [O/Fe] values of the order 
of +0.4\,dex, whereas our four LMC giants with low Al have [O/Fe]=+0.15 
to +0.20\,dex.  If confirmed on larger samples, this [O/Fe] difference 
between LMC and Galactic clusters could be attributed to a difference 
in the chemical evolution of the two galaxies.

\section{Evolution of metallicity with age in the LMC}

\subsection{New estimations of cluster ages}

Many papers have recently discussed the ages of the old 
Magellanic clusters compared to their Galactic counterparts
(Johnson el al. \cite{johnson 99}, Olsen et al. \cite{olsen 98},
Brocato et al. \cite{brocato 96}), coming to the common conclusion that 
Galactic and  Magellanic clusters could have been 
drawn from the same parent population. However, all age
determination methods are dependent (through direct or less direct ways)
on the assumed chemical composition of the clusters (overall
metallicities but also $\alpha$-element abundances).
We are now for
the first time, in a position to improve dramatically this knowledge
with precise abundances of individual cluster giants. 

ESO\,121-SCO3 and NGC\,1866 were found to have the same metallicity as 
previous estimations and, as a consequence, no effect on the age 
of these clusters is expected.

For NGC\,2210, on the other hand, our [Fe/H] determination is
+0.25\,dex higher than the value 
generally assumed to determine the age of the cluster. Age
determination by isochrone turnoff fitting is sensitive to the adopted
metallicity, an increase of 0.25\,dex inducing up to 2\,Gyrs younger 
ages (cf Figure 10 of Johnson et al. \cite{johnson 99}).  
But we found on the other hand, that NGC\,2210 
is oxygen poor compared to similar Galactic 
clusters, by $\sim -$0.2\,dex.
In this case, the opposite effect on the isochrones locus of a higher
iron and lower $\alpha$-element content cancels out almost exactly, 
so that the age of NGC\,2210 should remain very similar to the age of
the old Galactic clusters.
 
NGC\,1978 was found 0.54\,dex more metal deficient than
previous estimations, but the stars we have studied are rather cool 
and this result is more uncertain.  If this metallicity is confirmed,
the effect on age would be of a few 
tenths of billion years, making the cluster $\sim$10-20\% older than 
its current 2.2\,Gyrs estimate.

\subsection{Age-metallicity relation}

\begin{figure}
\resizebox{\hsize}{!}{\includegraphics{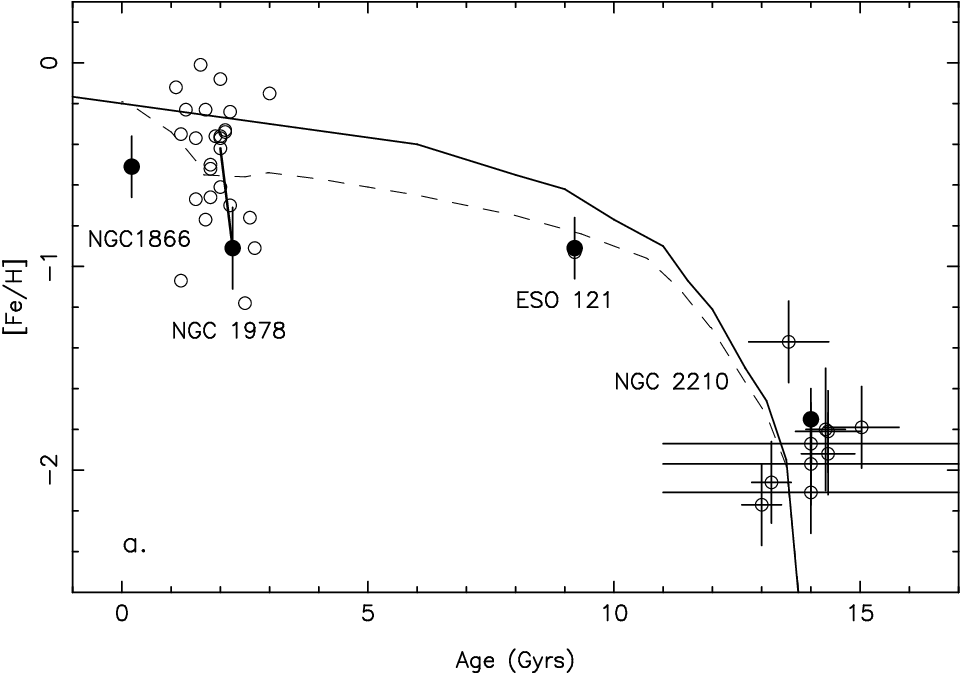}}
\resizebox{\hsize}{!}{\includegraphics{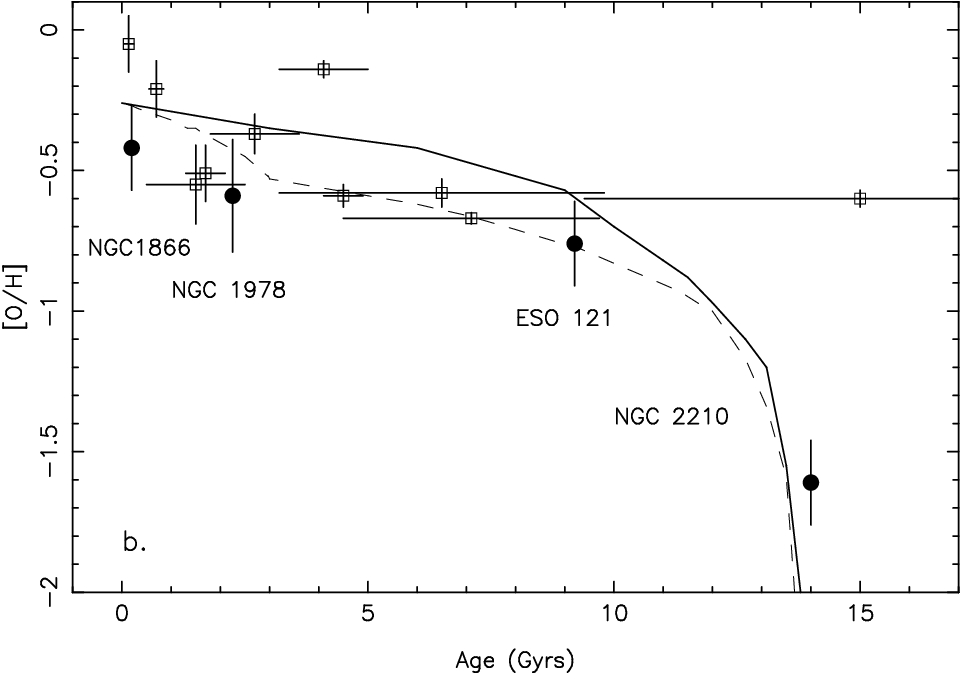}}
\caption{a- Age-metallicity relation derived from the four clusters in 
this paper, together with the data for LMC clusters from 
Geisler et al. (\cite{geisler 97}). 
Overplotted is a theoretical age-metallicity 
relation from a model using a continuous (thin line) and a bursting
(dashed line) SFR (Pagel \& Tautvai\v{s}iene \cite{pagel 98}). The same
clusters from the two sources are joined by thick lines.
b- Age-abundance for oxygen in the LMC, from the four clusters in 
this paper, together with data for PN from Dopita et
al. (\cite{dopita 97}). Curves as in a-.}\label{F-agemet}
\end{figure}

\begin{figure*}
\resizebox{\hsize}{!}{\includegraphics{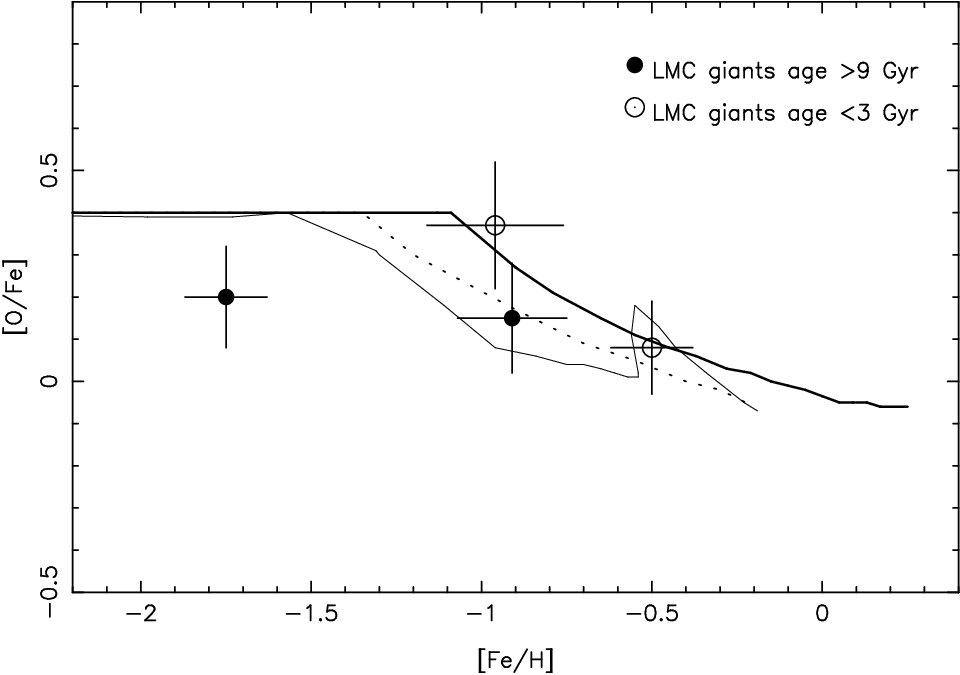}
	              \includegraphics{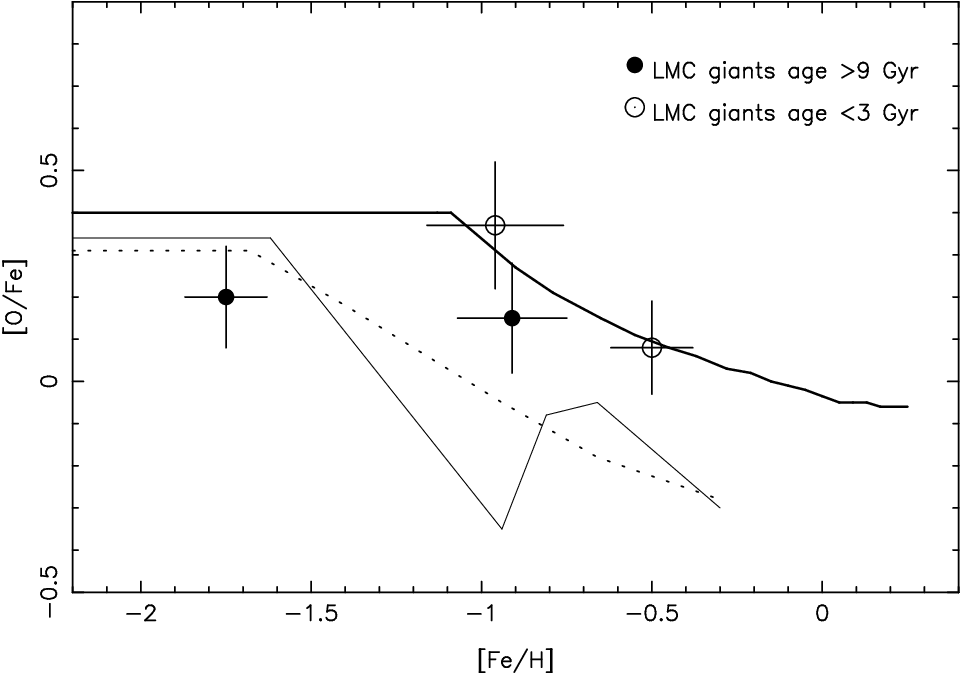}}
\caption{Evolution of the oxygen over iron ratios with metallicity in
the LMC.  
The mean oxygen abundance for the LMC clusters (this paper) 
are plotted together with
LMC chemical evolution models using a bursting
SFR (thin line) and a continuous SFR (dotted line) from: left panel
Pagel \& Tautvai\v{s}iene (\cite{pagel 98}) and right panel Tsujimoto et
al. (\cite{tsujimoto 95}). The thick line represents the
behavior of [O/Fe] in our Galaxy (Pagel \& Tautvai\v{s}iene \cite{pagel
95}).
}
\label{F-osfe}  
\end{figure*}

The LMC cluster distribution
has a long-known peculiarity: both metallicity and age distributions
are bimodal, with a well defined {\it gap} between 3-4\,Gyrs and
10-12\,Gyrs, corresponding to a minimum of the metallicity distribution
between $\sim -$1.0 and $\sim -$1.5\,dex. This bimodality has been
interpreted as the signature of two bursts of star formation, an
early one giving birth to the 12-15\,Gyrs clusters and a more recent
one some 3\,Gyrs ago, possibly triggered by tidal interaction of the
LMC with our Milky Way. 
Although the validity of such a conclusion on the global history of
the LMC is still a matter of debate
(is the cluster formation directly linked to the global star
formation ? what is the current rate of cluster disruption in the LMC
and what was it in the past ?), it is of great interest to investigate
whether hints of this bimodal star formation can also be found in the
age-metallicity relation.
Both clusters (see for example Geisler et
al. \cite{geisler 97} and Olszewski et al. \cite{olszewski 91}) 
and planetary nebulae (Dopita et al. \cite{dopita 97}) have been
used for this purpose, clusters having more precise age
determinations, while planetary nebulae have more accurate chemical
composition determinations (O, Ne, Ar, S).

In Fig.\ref{F-agemet}, the newly derived abundances of the four 
clusters studied here are overplotted onto the cluster sample of 
Geisler et al.  (\cite{geisler 97}) (panel a., metallicity is traced 
by [Fe/H]) and the planetary nebulae (PN) sample of Dopita et al.  
(\cite{dopita 97}) (panel b.  where metallicity is traced by 
[$\alpha$/H]).  
The ages of the old ($>$10\,Gyrs) clusters were not taken from Geisler
et al. \cite{geisler 97}, but adopted from more precise
determinations using deep CMDs (Brocato et al. \cite{brocato 96}, Olsen et al.
\cite{olsen 98} and Johnson et al. \cite{johnson 99}) and brought to
the same scale using an age of
14\,Gyrs for the Galactic comparison clusters.
Overplotted are predictions from Pagel 
\& Tautvai\v{s}iene (\cite{pagel 98}) semi-empirical model for two 
different star formation regimes: the full line is a continuous star 
formation rate, whereas the dashed line arise from two strong star 
formation episodes (14 and 3 Gyrs ago respectively) separated by a low 
star formation period.  In panel b., the abundances plotted are, in 
the case of PN, a mean of Ne, S and Ar, and for the four program 
clusters, the O abundance.  In both cases, as first claimed by Dopita 
et al.  (\cite{dopita 97}), the age-metallicity distribution seems to 
be compatible with a burst of star formation some 3\,Gyrs ago, 
triggering an increase of metal-abundances by a factor $\sim$3 around 
this age.  Of course, our sample is too small to be conclusive by 
itself, but the confirmation that there indeed exist clusters 
(NGC\,1978) as young as 2\,Gyrs and as metal-poor as clusters of ages 
9-10\,Gyrs (ESO\,121-SCO3) is by itself an interesting conclusion.

\section{Oxygen over iron ratios and chemical evolution}\label{oxygen}

The most powerful tool available to study the chemical 
evolution of a galaxy is to follow the behaviour, along time, of the 
abundance of elements which trace various nucleosynthetic channels and
sites. In particular, the evolution of elements produced in 
different mass-range progenitor stars, gives insight on the IMF and the 
SFR of the parent galaxy. The most well known such ratio is the
[O/Fe], which evolution along time (or metallicity) allows to
constrain the ratio of the number of massive supernovae (SNII) over
supernovae type Ia (SNIa). 
Up to recently the detailed study of the chemical composition of  
Magellanic objects were mainly restricted to the brightest stars, 
namely the young supergiants (cf e.g. Luck et al. \cite{luck
98}, Hill et al. \cite{hill 95}) and to the H\,II regions, 
which trace matter younger than $\simeq$0.1 Gyr. We are now in the
position to {\it follow along time}, the evolution of element ratios.

%
In the hypothesis that oxygen has not been depleted in the four old LMC
cluster giants (all except NGC2210\,B4364, see section \ref{mixing}), 
we can now  understand
the chemical evolution implications of the observed [O/Fe] ratios.  

Fig. \ref{F-osfe} displays the observed [O/Fe] versus [Fe/H] locus of
our LMC sample, compared with the 
predictions of two families of models computed for the LMC (see
caption for details). On the
left panel is the semi-empirical model by Pagel \& Tautvai\v{s}iene 
(\cite{pagel 98}),
which use the same IMF and yields than in the solar neighbourhood (SN), a
star formation rate (SFR) proportional to gas content, a gradual
inflow of unprocessed material and galactic wind proportional to the
SFR. Their prescriptions include a [O/Fe] ratio at present time such
that [O/Fe]$_{\rm LMC}$=[O/Fe]$_{\rm SN}$ at present time. Two models are
considered: one where the SFR is continuous, and the other where two
bursts occur 14\,Gyrs and 3~Gyrs ago respectively.
On the other hand, on the right panel of Fig. \ref{F-osfe} are models
by Tsujimoto et al. (\cite{tsujimoto 95}) which, at variance with the
Pagel \& Tautvai\v{s}iene (\cite{pagel 98}), allow the IMF slope to
change such that the [O/Fe]$_{\rm LMC}$=$-$0.2 at present time. 
\footnote{This major difference between the two families of models has been
triggered by a diverging interpretation of the observed LMC supergiants oxygen
abundances. A general agreement between various sources (cf Luck et
al. \cite{luck 98}, Hill et al. \cite{hill 97}) has found
[O/Fe]$_{\rm LMC}$=$-$0.2 from F-K supergiants, but, as argued by several authors 
(Luck \& Lambert \cite{luck lambert 92}, Hill et al. \cite{hill 97}, 
Pagel \& Tautvai\v{s}iene \cite{pagel 98}), 
this value should not be taken as absolute, but in
reference to stars of similar type in the SN, which also give
[O/Fe]$_{\rm LMC}$=$-$0.2 to $-$0.3\,dex, so that one can safely assume that
[O/Fe]$_{\rm LMC}$=[O/Fe]$_{\rm SN}$ at present time.}
The IMF needed to fulfill this constraint (where [O/Fe]$_{\rm LMC}
\neq$[O/Fe]$_{\rm SN}$) is steeper in the LMC than in the SN (in both
continuous and bursting SFR models).

Interestingly, none of the two families of models appear to be able to
reproduce the observed [O/Fe] of LMC clusters: while Tsujimoto et
al. (\cite{tsujimoto 95}) can almost reproduce the low [O/Fe] of the 
metal-poor cluster (thanks to their steeper IMF), they cannot reproduce
the relatively high [O/Fe] of the younger clusters at all;  Pagel \&
Tautvai\v{s}iene (\cite{pagel 98}) on the other hand, predict
compatible [O/Fe] in the young population, but too high [O/Fe] for the
older clusters.

The main lesson which can be learned from Fig. \ref{F-osfe} is that, if
the low [O/Fe] ratio in the older LMC clusters is confirmed, then the
[O/Fe] run with increasing metallicity would be extremely flat. This is a
hint that the LMC chemical evolution might have been driven by a
different fraction of SN\,II/SN\,I than in our own galaxy, SN~I
contributing a larger fraction of the iron in the LMC than in the solar neighbourhood.
This finding is still speculative at this point, as
we definitely need larger samples, both of globular cluster and old field
giants in the LMC to demonstrate it.
Also, extending this work to the SMC would be a complementary approach
to understand the role of the parent galaxy morphology on chemical evolution.

\begin{acknowledgements}
We thank the Science Operations on
Paranal and the UVES Science Verification team 
for the conduction of the observations and the timely public release of the
data to ESO member-states. 
\end{acknowledgements}

\end{document}